\documentclass[useAMS,usenatbib,usegraphicx]{mn2e}

\title[Ultra-fine structure in dark matter halos]{Modelling ultra-fine structure in dark matter halos}
\author[Fantin et al.]
{Daniele S. M.\ Fantin$^1$\thanks{E-mail: ppxdf@nottingham.ac.uk},
Michael R. Merrifield$^1$,
Anne M. Green$^1$\\
$^1$School of Physics \& Astronomy, University of Nottingham, 
    University Park, Nottingham, NG7 2RD\\
}
\begin{document}

\pagerange{\pageref{firstpage}--\pageref{lastpage}} \pubyear{2007}

\maketitle

\label{firstpage}

\begin{abstract}

  Various laboratory-based experiments are underway attempting to
  detect dark matter directly. The event rates and detailed signals
  expected in these experiments depend on the dark matter phase space
  distribution on sub-milliparsec scales. These scales are many orders
  of magnitude smaller than those that can be resolved by conventional
  N-body simulations, so one cannot hope to use such tools to
  investigate the effect of mergers in the history of the Milky Way on
  the detailed phase-space structure probed by the current
  experiments.  In this paper we present an alternative approach to
  investigating the results of such mergers, by studying a simplified
  model for a merger of a sub-halo with a larger parent halo.  With an
  appropriate choice of parent halo potential, the evolution of
  material from the sub-halo can be expressed analytically in
  action-angle variables, so it is possible to obtain its entire orbit
  history very rapidly without numerical integration.  Furthermore by
  evolving backwards in time, we can obtain arbitrarily-high spatial
  resolution for the current velocity distribution at a fixed point.
  Although this model cannot provide a detailed quantitative
  comparison with the Milky Way, its properties are sufficiently
  generic that it offers qualitative insight into the expected
  structure arising from a merger at a resolution that cannot be
  approached with full numerical simulations. Preliminary results
  indicate that the velocity-space distribution of dark matter
  particles remains characterized by discrete and well-defined peaks
  over an extended period of time, both for single and multi-merging
  systems, in contrast to the simple smooth velocity distributions
  sometimes assumed in predicting laboratory experiment detection
  rates.  In principle, this structure contains a wealth of
  information about the formation history of the Milky Way's dark
  halo. 

\end{abstract}


\section{Introduction}
Dark matter (hereafter DM) appears to be the dominant mass component
of galaxies and large-scale structures in the Universe. The first
evidence came in the 1930s (Zwicky 1933, Smith 1936), but it was only
in the 1970s that observations of the rotation curves of galaxies
demonstrated that DM dominates the masses of galaxies (Rubin \& Ford
1970, Rubin, Thonnard \& Ford Jr 1980). These observations showed that
many rotation curves are approximately flat, or even rising, in the
outer region of galaxies, where there is little luminous matter and so
a Keplerian decline is expected.  Subsequently, work on the
hierarchical structure formation paradigm showed that non-baryonic
material known as ``cold dark matter'' (CDM) is required to match the
observed large-scale structure of the Universe (Peebles 1982). The
term ``cold'' derives from the fact that this material was
non-relativistic at the epoch of matter-radiation equality. The
density of this CDM has subsequently been indirectly measured by
various experiments such as 2dFGRS (Percival et al., 2001), WMAP
(Dunkley et al. 2008) and the Sloan Digital Sky Survey (Tegmark et
al. 2006).

Particle physics provides us with various well-motivated candidates
for the CDM, including weakly interacting massive particles
(WIMPs). WIMPs can potentially be directly detected in the laboratory
via their elastic scattering on target nuclei, and numerous
experiments are currently underway to try to detect this phenomenon
(e.g. Angle et al. 2007, Ahmed et al. 2008). The signals expected in
these experiments, the number of recoil events per unit energy (and in
some case its temporal and angular dependence), depend on the WIMP
velocity distribution in the solar neighbourhood. A single stream of
dark matter particles produces a step in the energy spectrum,
detectable by a detector. The energy at which the step
occurs is determined by the speed of the particles composing the
stream (in the rest frame of the detector), while the height and
position of the step vary annually, due to the Earth's orbit.
Multiple streams would lead to a more complicated picture, and a
superposition of enough such streams would ultimately be
indistinguishable from a smooth distribution, depending on the
detector resolution.  The question that we seek to address here is
what form one might expect for this distribution in reality.

Predictions of the expected signals are often based on simplified
models, which assume, for example, that the WIMP velocity distribution
is Maxwellian (Freese, Frieman \& Gould 1988) or a multivariate Gaussian
(Evans, Carollo \& de Zeeuw 2000, Helmi, White \& Springel
2002). These models rely on the assumption that the Milky Way halo has
reached a steady state so that the ultra-local DM phase-space
distribution is smooth. However since structures form hierarchically,
and the age of the Universe is not large compared with relevant
dynamical timescales (such as the crossing time), this assumption is
somewhat questionable.

Numerous N-body simulations have been performed studying the
hierarchical formation and evolution of DM halos. Such simulations
find that DM halos contain ubiquitous substructure, in particular in
their outer regions (Moore et al.\ 1999, Klypin et al.\
1999). However, the salient question here is whether or not the DM
distribution is smooth on the scales probed by direct detection
experiments.  Unfortunately, the resolution of even the best N-body
simulations is many orders of magnitude larger than the relevant
scales. The Sun's circular velocity around the centre of the Galaxy is
$v_{\odot} \approx 200 \; \rm{km/sec}$, so that over a course of a
year a terrestrial DM detector travels a distance
\begin{equation}
\rm{r_{det}} \approx v_{\odot} \; \tau_{\rm{exp}} \sim (200 \;
\rm{km/sec}) \; ({1 \; \rm{yr}}) \sim 0.1 \; \rm{mpc} \,,
\end{equation} 
whereas current N-body simulations cannot resolve scales smaller than
${\sim \; 100 \, \rm pc}$ (e.g. Diemand et al. 2007).

This represents an insurmountable problem for the conventional
simulation techniques, indicating that a completely different,
specialized approach is necessary to describe in detail the ultra-fine
DM distribution probed by direct detection experiments. A first
attempt at such a specialized simulation was carried out by Stiff and
Widrow (2003; hereafter SW). Their method used a reverse simulation
process to calculate the DM speed distribution, $f(v)$, at a single
spatial point of the phase space, representing a detector. More
specifically they ran a simulation of the formation of a DM halo and
at the end of the simulation put down a uniform grid (in velocity
space) of mass-less test particles at the point of interest. They then
evolved the test and simulation particles back to the initial time,
found where the phase-space sheet of the test particles intersects the
initial DM phase-space distribution, and hence calculated the density
of the test particles at the final detector position. In this way,
they found that the DM distribution in the solar neighbourhood is
characterized by a number of discrete peaks. This numerical approach
allowed them to use initial conditions which reproduce a realistic
hierarchical-formation model for the Milky Way. However, a drawback of
this technique was that it proved numerically unstable, and, in order
to stabilize the reverse integration, SW were obliged to introduce a
softening length of some $20\, {\rm kpc}$ into the gravitational force
law that they applied. This softening is worryingly large as it
significantly exceeds the solar radius in the Milky Way, $R_0 \approx
8.5\, {\rm kpc}$, so might be expected to affect the inferred DM phase
space distribution impinging on a terrestrial detector.

More recently Vogelsberger et al. (2008) have formulated a technique
for calculating the evolution of the fine-grained DM density in both
static potentials and N-body simulations. They argue that the
small-scale DM distribution that a terrestrial experiment would
observe can be described by a multivariate Gaussian, in apparent
contradiction to the SW results, perhaps indicating that the SW
analysis had been compromised by the modification that they had to
make to the gravitational force.

In this paper we develop a complementary approach to the analysis of
the ultra-fine DM distribution. Following SW, we calculate the DM
distribution in the solar neighbourhood via a backward evolution
method, but using a simplified model for the potential that allows the
system to be expressed in action-angle (hereafter AA)
variables. Although this simplified potential provides a less
realistic representation of the Milky Way, its qualitative properties
are similar, and it has the great benefit of being analytically
soluble. Thus, the gravitational force does not have to be
artificially softened (allowing one to test whether this effect did compromise the
SW results), and one can very rapidly explore parameter space without
the computational overhead of numerical integration.

The paper is organized as follows. In Section~2 we present the
simplified model that we use to describe the interaction between a
galaxy like the Milky Way and a merging sub-halo. Section~3 contains
our initial results, and we conclude in Section~4 with a discussion.

\section{The model}

\label{sec:model}

To study the evolution of the ultra-fine DM distribution contributed
by a merging halo in a massive galaxy like the Milky Way, we adopt the
isochrone potential,
\begin{equation} \label{eq:potential}
\Phi(r) = \frac {G M}{b + \sqrt{b^2 + r^2}}, 
\end{equation} 
for the potential of the massive system.  Although not intended as a
realistic model for a complex system like the Milky Way, it can be
tuned through its mass $M$ and characteristic lengthscale $b$ to
approximate a range of systems.  More importantly, orbits in such a
potential can be calculated analytically by expressing the dynamics in
AA variables, greatly simplifying the time evolution calculations
(McGill \& Binney 1990; Gerhard \& Saha 1991).

Specifically, the Hamiltonian can then be expressed in terms of the
actions, ${\bf J}$, as
\begin{equation} 
H(\mathbf{J}) = -\frac{2}{(2J_r + L + \sqrt{4+L^2})^2}, 
\end{equation}
where
\begin{equation} 
L \equiv \mid \mathbf{J}_l + \mathbf{l}_z \mid 
\end{equation}
is the magnitude of the angular momentum vector. The actions remain
constant with time, which we can express in spherical coordinates as

\begin{equation} 
J_{i}(t) =  J_{i}(t_{0}).
\end{equation}
where $i=r,l,\varphi$. The corresponding angle variables,
${\theta_i}$, can be determined from the solution to Hamilton's
equation,
\begin{equation}
\dot{\theta_{i}} = \frac{\partial{H}}{\partial{J_i}} \equiv \Omega_i(\mathbf{J}),
\end{equation} 
where the $\Omega_i$ are the corresponding angular frequencies. These
terms therefore evolve linearly with time, allowing the solution at
any epoch (either forward or backward in time) to be expressed
trivially in terms of the initial values,
\begin{equation} \label{eq:angles}
\theta_{i}(t) =  \theta_i(t_0) -  \Omega_{i}(t-t_{0}).
\end{equation}
More details, including analytic expressions relating the AA variables
(${\bf J}, {\theta}$) and Cartesian coordinates (${\bf x}, {\bf v}$)
can be found in McGill \& Binney (1990) and Gerhard \& Saha (1991).

In the present context, we are interested in the velocity distribution
of particles passing through a fixed position (representing a DM
detector) and its evolution with time. At this location, we can pick
any velocity and analytically evolve these phase space coordinates
backwards in time to determine their initial phase space location. We
can then calculate the amount of material in a merging satellite that
originated from these initial phase-space coordinates, and hence will
end up at the selected velocity in a terrestrial detector today. By
stepping through all such velocities, we can map out the present-day
phase-space structure of this disrupted merging satellite on an
arbitrarily fine spatial scale.

To complete this model, we need to specify the initial DM phase-space
distribution of the merging halo. The simplest representation that
provides enough freedom to explore the dependence on the properties of
this merging halo is provided by a bivariate Gaussian,
\begin{equation} \label{eq:disrt-func}
f(\mathbf{r},\mathbf{v})\propto \, e^{-[{(\mathbf{r}-\mathbf{r_0})^2}/{2h^2})]} \; 
e^{-[({\mathbf{v}-\mathbf{v_0})^2}/{2\sigma_{\rm{v}}^{2}}]} \ ,
\end{equation}
where $h$ models the initial spatial extent of the merging halo while
$\sigma_{v}$ defines its velocity dispersion. We also impose the
physically-motivated limit $v< v_{\rm esc}$ where $v_{\rm esc}$ is the
escape velocity of the parent galaxy,
\begin{equation} 
v_{\rm{esc}} = \sqrt{\frac{2k}{b+\sqrt{b^2+r^2}}}.
\end{equation}

In summary the principal steps in constructing this model for the present-day fine-scale phase-space distribution due to a merging satellite halo are:
\begin{enumerate}

\item Select the spatial and velocity scale for the satellite and its initial position and velocity, and the time in the past, $t_0$, at which it was at that location.

\item Choose the present-day phase space coordinates of the detector location and a particular velocity ${\bf v}$.

\item Transform these phase-space co-ordinates into AA variables.

\item Analytically evolve these AA coordinates back to $t_0$.

\item Transform them back into Cartesian co-ordinates.

\item Evaluate the initial phase space density due to the merging satellite for this phase-space location via
equation~~(\ref{eq:disrt-func}), which is then also the phase space density in the present-day detector at velocity ${\bf v}$.

\item Repeat for a grid of velocities at this location to map out the full velocity distribution within the detector.

\end{enumerate}
Clearly, this backwards-in-time approach allows us to pinpoint
efficiently those DM particles from the initial merging satellite that
are to be found passing through an arbitrarily-small detector today,
and by choosing the grid of velocities appropriately we can map out
the velocity structure with any resolution that we desire.

\section{Results}

%
%

\begin{figure}
\includegraphics[height=0.23\textheight, width=0.43\textwidth]{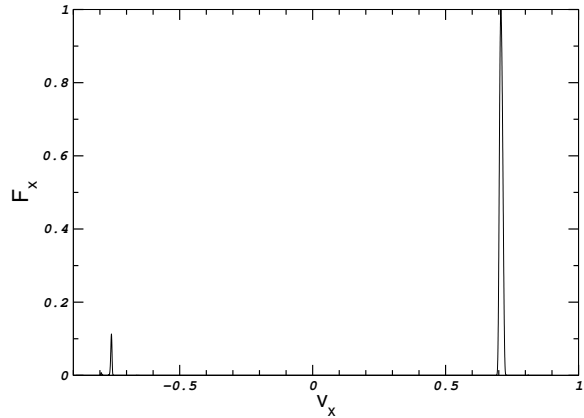}
\caption{The distribution function of the component of the speed in the direction of the merger, denoted as $x$, at the fixed position $\mathbf{r}=(1,0,0)$ at time $t=90$ (when the satellite is on its second orbit, corresponding to $\sim 1.20\,{\rm Gyr}$ for a Milky-Way like galaxy).}
\vspace{0.5cm}
\label{fig:mycode90}
\end{figure}

%
%

\begin{figure} 
\includegraphics[height=0.23\textheight, width=0.43\textwidth]{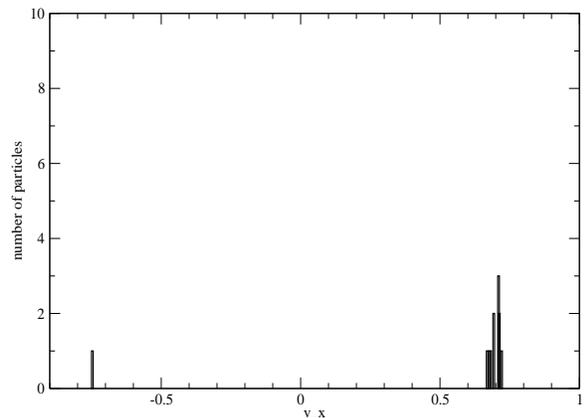}
\caption{Configuration as for Fig.~\ref{fig:mycode90}, but with the evolution of the satellite calculated by evolving the co-ordinates of $10^{5}$ particles forwards in time.} 
\label{fig:Kuijken90}
\end{figure}

In this section we present some illustrative results obtained using
the technique described in Sec.~\ref{sec:model}. This analysis is not
exhaustive, nor is it intended to be used to describe quantitatively
the merger history of the Milky Way. However, the control that we can
impose on this simplified model, and the speed at which it can be
computed at arbitrarily high resolution, means that it can be used to
obtain unique qualitative insights into the likely signature of a halo
merger event in a terrestrial DM detector.

As described above, this model simulates the fall of a DM halo into a
larger gravitational potential. The satellite is injected into the
host galaxy and we calculate how the velocity distribution of
particles passing through a fixed spatial position varies with
time. The particles composing the satellite initially have similar
(but not equal) energies. As a consequence, they follow slightly
different orbits, have different orbital periods and with time they
spread out in physical space.

As a concrete example we consider a satellite initially at position
${\bf r}_{0} = (-5, \, 0, \, 0)$, with velocity ${\bf v}_{0}= (0, \,
0.05, \, 0.1)$ and initial phase-space distribution function given by
eq.~(\ref{eq:disrt-func}). The spatial and velocity dispersions are
respectively $h=0.5$ and $\sigma_{v}=0.1$ (scaled to Milky Way units
corresponding to $ \sigma_{v} \sim 60 ~ \rm{km~s^{-1}}$). We use the
reverse AA evolution technique to calculate the velocity distribution
at a position ${\bf R}=(1,0,0)$ at a range of times, which we can
readily scale into physical ``Milky Way'' units corresponding to the
detector's location at $R_ 0 \sim 8.5\,{\rm kpc}$. This configuration
has been chosen to describe a cold and concentrated satellite (as
expected in reality), merging into the parent galaxy along $x$, the
axis defined by the direction of the merger, from a large distance
(corresponding to ${\bf r}_{0} \sim 40 \, {\rm Mpc}$).

In Figure~\ref{fig:mycode90} we plot the distribution of the
$x$-component of the velocity at the solar radius at time $t=90$. For
a Milky-Way-like parent galaxy, this corresponds to a physical time
$t=1.2 \, {\rm Gyr}$. At this stage of the merger the satellite is on
its second orbit. The satellite is still fairly coherent, but it is
beginning to be disrupted by tidal forces. The large peak at positive
speed corresponds to particles which are on their second orbit, while
the smaller peak at negative speed is part of a tidal tail which has
not yet finished its first orbit.

\begin{figure} 
\includegraphics[height=0.23\textheight, width=0.43\textwidth]{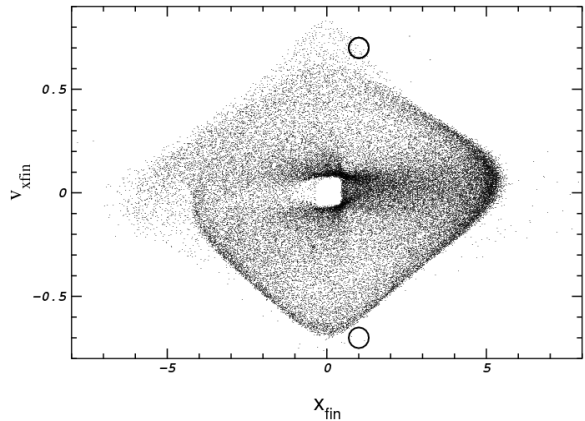}
\caption{Configuration of the satellite in the x-component of the phase-space at $t=90$ using the initial conditions of Fig.~\ref{fig:mycode90}. Two black circles pinpoint the regions the peaks of Fig. \ref{fig:Kuijken90} originate.}
\label{fig:xf-vxf90}
\end{figure}

To illustrate the power of the reverse AA evolution technique, we also
carried out a simulation of the same configuration using the more
conventional approach of evolving $10^{5}$ DM particles initially in
the satellite halo forward in time. Figure~\ref{fig:Kuijken90} shows a
histogram of the speed in the x-direction of the particles at the same
fixed point at the same time as for the backwards evolution
calculation in Fig.~\ref{fig:mycode90}. Clearly, the same peaks are
reproduced, but since the vast majority of particles in this forward
evolution ends up nowhere near the Earth, the sampling of these peaks
is extremely poor. This problem becomes even more severe at later
times as the particles populate a larger volume of
phase-space. However, the forward evolution does provide complementary
information by giving the big picture of how the satellite is being
disrupted. Figure~\ref{fig:xf-vxf90} shows the ($x,v_{\rm x}$) phase
space co-ordinates of $10^{5}$ particles evolved forwards, showing how
they have already spread out in phase space, with the peaks in the
terrestrial detector arising from particles that have completed one or
two orbits. The regions from which those particles come are labelled
in the plot with circles.

\begin{figure*}
\centering
\includegraphics[width=\textwidth]{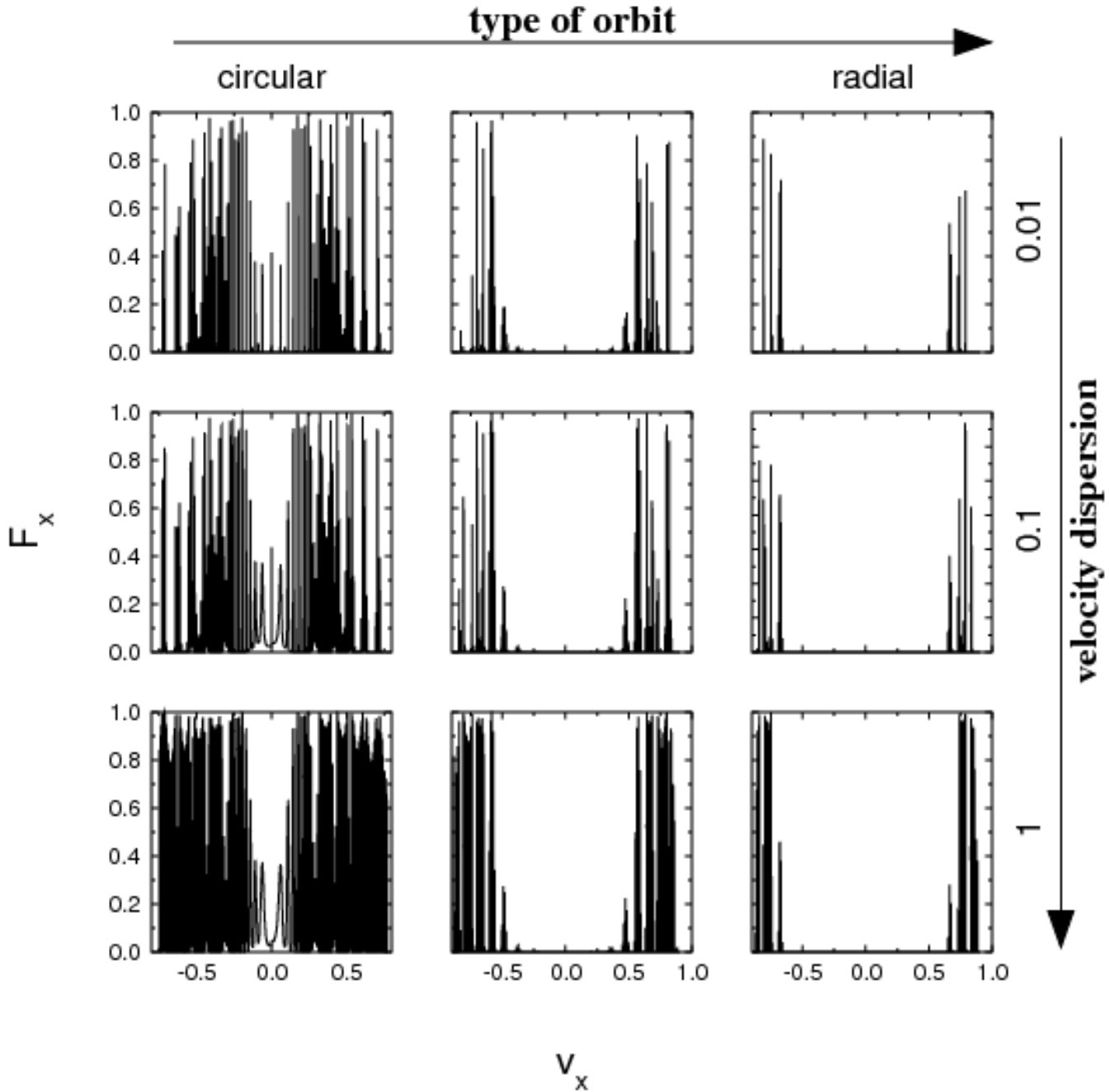}
\caption{The distribution function of the component of the speed in
  the direction of the merger as a function of the velocity dispersion
  of the halo and of the orbit performed in the galaxy potential. The
  nine snapshots depict the system at $t \sim 14\,{\rm Gyr}$. The
  orbit of the DM halo is taken into account moving from the left
  (circular orbit) to the right (radial). On the y-axis the result due
  to a change caused by an increase (from the top to the bottom) of
  the velocity dispersion is plotted.}
\label{fig:multiplot}
\end{figure*}

A further indication of the flexibility of the backwards approach is
provided by Figure~\ref{fig:multiplot}. Each panel depicts the
simulated phase-space distribution in a terrestrial detector after $t
\sim 14\,{\rm Gyr}$ for a different set of initial conditions that
vary the orbit of the satellite and its internal velocity
dispersion. Because of the analytic nature of the orbit integration,
this calculation was as simple as the shorter-evolution illustration,
with no loss of precision in the results.  As this figure illustrates,
the velocity distribution does depend on the merging halo parameters:
the higher velocity dispersion sub-halos, for example, wrap around the
Milky Way more quickly, creating more peaks in the distribution.
Nonetheless, the presence of persistent discrete fine phase space
structure is a generic property of all the mergers.

%
%

\begin{figure} 
\includegraphics[height=0.23\textheight, width=0.43\textwidth]{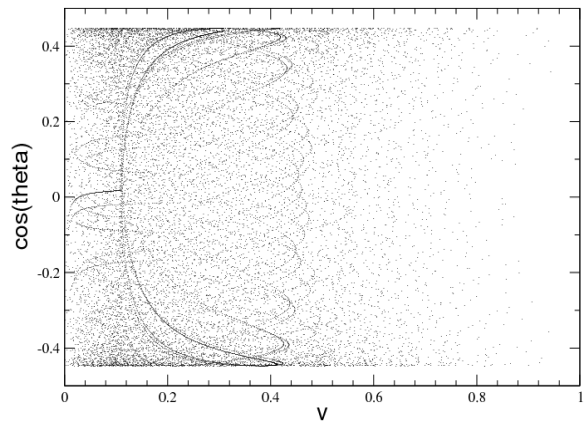}
\caption{ The speed and angle with respect to the direction of Solar
  motion of the particles impinging on a terrestrial DM detector from
  the merger described in the text after a time $t \sim 14\,{\rm
    Gyr}$.}
\label{fig:speed-pop992}
\end{figure}

This simplified model is not intended to predict quantitatively the
experimental signal that terrestrial DM detectors should see, but we
can take the qualitative analysis one step closer to the laboratory by
considering the physical quantities that are most relevant for such
experiments, particularly those with directional sensitivity. For such
detectors, a useful diagnostic is provided by the speed of the DM
particles as a function of the angle at which they impinge on the
detector (measured relative to the direction of Solar motion in the
Milky Way). Figure~\ref{fig:speed-pop992} shows this plot for the
simulation described above at time $t=14\,{\rm Gyr}$. Although the DM
particles are quite well spread through the parameter space, it is
apparent that even at this late time there is significant structure
apparent in the plot, which would have an impact on the detectability
of this particular merging sub-halo, and might even ultimately be used
to reconstruct its origins.

%
%

\begin{figure} 
\includegraphics[height=0.23\textheight, width=0.43\textwidth]{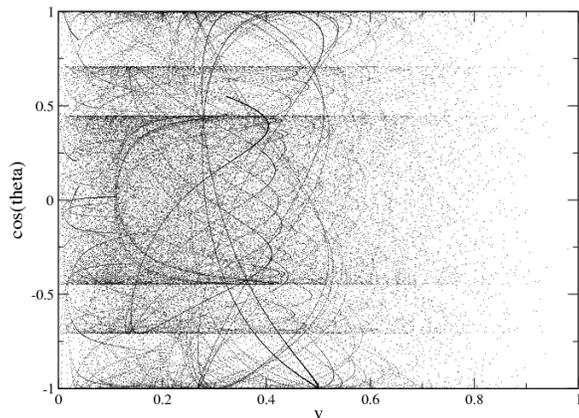}
\caption{The DM speed versus angle plot, as described in
  Fig.~\ref{fig:speed-pop992}, showing the consequences of three
  sub-halos merging into a Milky-Way-like potential well from a range
  of distances and directions.}
\label{fig:multisignal}
\end{figure}

In reality, of course, the Galactic halo is made up of multiple
merging events, so the velocity distribution should be even more
complex than that shown in Fig.~\ref{fig:speed-pop992}.  Since to a
good approximation these multiple mergers do not interact with each
other, we can model such a sequence by simply adding results like
those obtained in Fig.~\ref{fig:speed-pop992}.  As
Fig.~\ref{fig:multisignal} confirms, such superpositions further
complicate the structure, but certainly do not produce a simple smooth
distribution.  A detector with even relatively crude angular
resolution would be able to pick out the strong horizontal structures
in this diagnostic figure, which arise from the constraint imposed by
the cut-off at escape speed for DM particles from each individual
merger event, suggesting that much information might be gleaned about
the merger history of the Milky Way halo from future terrestrial DM
detectors.
%
%

\section {Conclusions}

As a simple model for the signal that might be detected in a
terrestrial DM detector due to a single sub-halo merger, we have
studied the distribution of particles resulting from the merger of a
satellite into a parent galaxy described by an isochrone
potential. Although not intended as a quantitative match to the Milky
Way, this form of potential has the great benefit that the dynamics of
the merger can be calculated remarkably simply, and the satellite can
be evolved forwards or backwards in time analytically using
action-angle variables. This simplicity allows us to quickly and
accurately calculate the velocity distribution at any time at
arbitrarily high spatial resolution, which is vital if we want to
understand the sub-mpc-scale structure of the Milky Way's halo probed
by terrestrial DM detectors. We find that, even at late times (up to
$14\,{\rm Gyr}$) when the particles have spread out through
phase-space, the velocity-space distribution function is characterized
by discrete and distributed peaks. In agreement with Stiff \& Widrow
(2003), we find that the parameters that dictate the detectability of
DM, particularly for experiments with directional sensitivity, contain
persistent significant structure imprinted by the original merging
sub-halo, which could well impact on the detectability of DM, as well
as shedding light on the properties of the progenitor sub-halo. 

Although the situation becomes more complex when one considers a halo
built up from multiple mergers over the lifetime of the Galaxy, the
evidence suggests that significant amounts of fine-grained features
persist.  Although such structures may have an impact on detectors'
ability to make an unequivocal detection of the DM halo, they also
raise the fascinating possibility that it may be possible to use them
to unravel the complete merger history of the Milky Way.

\section{Acknowledgments}

MRM and AMG gratefully acknowledge support from STFC fellowships.


\label{lastpage}

\end{document}